\documentclass[a4paper,11pt]{article}
\pdfoutput=1 %
\usepackage{jcappub} 

\usepackage[T1]{fontenc} % if needed

\usepackage[utf8]{inputenc}
\usepackage[dvipsnames]{xcolor}
\usepackage{hyperref} 
\usepackage{url}
\usepackage{graphicx}
\usepackage{esvect}
\usepackage{array}
\usepackage{diagbox}
\usepackage{graphics}
\usepackage{subcaption}
\usepackage{natbib}
\usepackage{mathrsfs}
\usepackage{blindtext}
\usepackage{comment} %For various lines comments
\usepackage{placeins} %For the \FloatBarrier command
\usepackage{multirow}

\usepackage[normalem]{ulem}
\usepackage{color}

\usepackage{soul}

\usepackage{supertabular}
\usepackage{tabularx}

\title{Impact of inhomogeneous diffusion on secondary cosmic ray and antiproton local spectra} %radially-dependent

\author[a, b]{Álvaro Tovar-Pardo}
\emailAdd{atovarpardo@gmail.com}
\author[a, b, c]{Pedro~De~La~Torre~Luque}
\emailAdd{pedro.delatorre@uam.es}
\author[a, b]{Miguel A. Sánchez-Conde}
\emailAdd{miguel.sanchezconde@uam.es}

\affiliation[a]{Instituto de F\'isica Te\'orica UAM-CSIC, Universidad Aut\'onoma de Madrid, C/ Nicol\'as Cabrera, 13-15, 28049 Madrid, Spain}
\affiliation[b]{Departamento de F\'isica Te\'orica, M-15, Universidad Aut\'onoma de Madrid, E-28049 Madrid, Spain}
\affiliation[c]{The Oskar Klein Centre, Department of Physics, Stockholm University, AlbaNova\\
  SE-10691 Stockholm, Sweden}
\date{\today}

\abstract{Recent $\gamma$-ray and neutrino observations seem to favor the consideration of non-uniform diffusion of cosmic rays (CRs) throughout the Galaxy. 
In this study, we investigate the consequences of spatially-dependent inhomogeneous propagation of CRs on the fluxes of secondary CRs and antiprotons detected at Earth. A comparison is made among different scenarios in search of potential features that may guide us toward favoring one over another in the near future. We also examine both the influence of inhomogeneous propagation in the production of secondary CRs from interactions with the gas, and the effects of this scenario on the local fluxes of antiprotons and light antinuclei produced as final products of dark matter annihilation. Our results indicate that the consideration of an inhomogeneous diffusion model could improve the compatibility of the predicted local antiproton flux with that of B, Be and Li, assuming only secondary origin of these particles. In addition, our model predicts a slightly harder local antiproton spectrum, making it more compatible with the high energy measurements of AMS-02. 
Finally, no significant changes are expected in the predicted local flux of antiprotons and antinuclei produced from dark matter among the different considered propagation scenarios.

}

\begin{document}
\maketitle
%\flushbottom

\section{Introduction}
\label{sec:intro}

Over the last century, cosmic rays (CRs) have proven to be a powerful tool for the study of multiple areas in physics, from Galactic astrophysics~\cite{Widmark_2023, ruszkowski2023cosmic, Semenov_2021} to high-energy physics \cite{1986ApJ...300..474F, 1991crpp.book.....G}. Foremost, the indirect search for dark matter (DM) through the detection of CRs has recently emerged as a promising field~\cite{Conrad_2017, DONATO201441}. Additionally, the current interest shown by the scientific community in the field has increased in recent years thanks to the data provided by modern detectors, mainly, the AMS-02 experiment \cite{AMS_gen}\footnote{\texttt{\hyperlink{https://ams02.space/}{https://ams02.space/}}}, which showcases an unprecedented level of precision in its data in the energy range of GeV-TeV per nucleon. %AMS-02's substantial contribution has propelled the research of acceleration mechanisms and transport models of CRs into the spotlight of astroparticle physicists \textcolor{red}{(rephrase?)}.

Accelerated mainly in Supernova Remnants (SNRs), primary CRs (charged particles produced in the Big Bang Nucleosynthesis or in the stellar interior) travel along the Galaxy being deflected by the Galactic magnetic fields, interacting acollisionally with turbulent plasma waves and producing secondary particles by spallation phenomena with the gas present in the interstellar medium (ISM). Primary and secondary CRs eventually reach the Earth, allowing us to study, among others, the diffusive models that characterize their transport~\cite{Strong_1998}, the Galactic environment or, ultimately, the nature of DM particles and the morphology of the structures they form~\cite{Ginz&Syr}.

The widely embraced scientific perspective holds that CRs primarily follow a diffusive pattern of motion along our Galaxy, stemming from their interactions with magnetohydrodynamical waves~\cite{Fornieri_2021, Blasi_2012}. The exchange of energy with these waves causes the reacceleration and deflection of a CR. All the diffusive behavior of CRs is enclosed in the so-called diffusion coefficient, which was found to follow a power law in rigidity with a spectral index $\delta$~\cite{maurin2002galactic, 2018AdSpR..62.2731A},
\begin{equation}
    D(\rho)\sim D_0 \beta ^\eta \left( \frac{\rho}{\rho_0} \right)^\delta
    \label{eq:1}
\end{equation}
where $\beta$ is the speed of the particles in units of speed of light and $\eta$ is an exponent fitted to the data.
Both the normalization constant, $D_0$, and the spectral index are values that have to be obtained experimentally. Measured spectra lead to typical local values of $\delta \approx$ 0.50 \cite{Blasi_2012}, and a normalization constant in the order of $D_{0} \sim (3-5)\times 10^{28}~cm^2s^{-1}$ at $\rho \sim$ 1 GV \cite{Strong_2007}.
%The way in which CRs diffuse throughout the Galaxy dictates vastly the amount of secondary particles arriving at Earth, and vice versa \textcolor{red}{(ref. to an Eq. $\phi_S/\phi_P \sim \sigma(E)D^{-1}?$)}.
The most widely used observable to determine the diffusion parameters is the secondary-to-primary flux ratio, which depends almost exclusively on the diffusion parameters and the effective cross-section of secondary production. Hence, cross section values play a fundamental role in the evaluation of the diffusion coefficient. Usually the high uncertainties associated with the effective cross-section affect both the determination of diffusion parameters and the identification of possible DM annihilation signals. For instance, cross section uncertainty associated with boron or antiproton production can be up to $20\%$~\cite{Doetinchem_2020, Luque_etal_2021, Genolini_2018}.

%\begin{equation}
%    \frac{\phi_S}{\phi_P} \sim \frac{\sigma(E)}{D(E)} 
%    \label{eq:ratio}
%\end{equation}

On top of this, it is unclear if the spectral behavior of the diffusion coefficient is equal in the whole Galaxy or if, on the contrary, it is only valid for our local neighbourhood. In other words, whether diffusion occurs uniformly or if it exhibits a spatial dependence is a fundamental open problem~\cite{Schroer, Dundovic_2020}. At present, most (if not all) analyses of CRs employ a diffusion coefficient that is uniform across the Galaxy. Nevertheless, there are indications suggesting that this may not necessarily be the case. The Milky Way exhibits non-homogeneous density of matter and magnetic fields, being proved that there are more structures and astrophysical activity in the Galactic center (GC)~\cite{Gabici:2019jvz}. Therefore, it is expected that the diffusion of CRs in the GC differs from that in our solar neighbourhood.
Notably, the Fermi Large Area Telescope (\textit{Fermi}-LAT), on board the NASA Fermi satellite~\cite{Atwood_2009} has performed observations indicating a discernible increase in the GeV $\gamma$-rays diffuse flux directed towards the GC~\cite{Ackermann_2017, DAYLAN20161, Calore_2015, Abazajian_2014, Hooper_2011}. Furthermore, recent data on $\gamma$-ray emissions at the PeV~\cite{TIBET, Cao_2023} and neutrino emissions at the TeV-PeV range, as disclosed by the IceCube experiment~\cite{IceCube}, have revealed an unexpectedly elevated flux. 
This phenomenon can be adequately elucidated by postulating that the propagation of CRs is not uniform across the Galaxy. For instance, recent studies~\cite{Luque_PeV, Gaggero_2015} has offered a natural explanation for the excesses found in the diffuse $\gamma$-ray emission by introducing a non-homogeneous diffusive model in CR transport, wherein a spatially-dependent spectral index is taken into account.

Additionally, this can be essential in searches of DM using CRs. The WIMP (Weakly Interactive Massive Particles) model posits the existence of DM particles capable of producing particle-antiparticle pairs with energies on the order of the GeV \cite{Roszkowski_2018, kamionkowski1997wimp}. 
%Hence, AMS-02 dataset have triggered the interest in the indirect search for DM through the detection of CRs. 
In this regard, the measurement of antiparticles, such as antiprotons and antinuclei, could provide clear signals of WIMP annihilation due to their very low backgrounds, e.g.,~\cite{Doetinchem_2020}. Finding DM signals requires a highly accurate model of the transport of CRs: a thorough understanding of CRs spectra is necessary for the subsequent detection of anomalies in the data.

This study explores the production of CRs in our Galaxy considering a spatially-dependent inhomogeneous diffusion model, and emphasizing its effect on the production of secondary CRs and antiprotons. The aim of this study is to demonstrate the impact that more realistic propagation setups have in the predicted fluxes of secondary CRs at Earth, yet we note that a full evaluation of the uncertainties related to inhomogeneous propagation setups is still far from being achievable.
%\footnote{Positron fluxes are not contemplated in this work since energy losses prevent their propagation over large distances, obtaining no differences in their spectra in both models \textcolor{red}{(ref?)}.}.
In Section \ref{sec:method}, we outline the main motivations for this work, providing a concise discussion of the CRs transport models and the considered setup. Section~\ref{sec:Sec} presents a comparison of the predicted local spectra of secondary CRs and antiprotons for the propagation scenarios considered. First, we examine a homogeneous transport model, treating it as the benchmark scenario for subsequent comparative studies with non-uniform models. Then, we analyze the results obtained with the inhomogeneous model from Ref.~\cite{Luque_PeV} and perform a combined fit to AMS-02 data. The results indicate that we can not distinguish between both scenarios with the current data, although inhomogeneous transport characterized by a spatially-dependent spectral index could alleviate some existing tensions. In Section~\ref{sec:DM}, we study the antiproton flux produced by DM annihilation for various WIMP masses and discuss the impact of neglecting the spatial dependence of the diffusion coefficient in current indirect DM searches. We discuss our findings and conclude in Section \ref{sec:conc}.

\section{Cosmic-ray propagation models}
\label{sec:analysis}

As discussed above, the validity of the local spectral shape of CRs in different parts of the Galaxy is a fundamental open problem that has been actively studied over the past decade. Both, theoretical expectations and recent data suggest that the diffusion coefficient is not constant throughout the Galaxy~\cite{Cao_2023, Schroer, TIBET, IceCube, Cerri_2017}.  In particular, a recent model~\cite{Luque_PeV, Gaggero_2015} implements a spectral index of the diffusion coefficient that depends on the distance from the GC to explain the hardening of the Galactic diffuse $\gamma$-ray flux observed by Fermi-LAT \cite{Fermi_Gamma, Lipari_2018}, and takes the following form:
\begin{equation}
    \delta(r) = \delta_0 \cdot r + \delta_1, 
    \label{eq:delta}
\end{equation}
where $\delta_0$, $\delta_1$ are parameters that can be determined through fits to the diffuse $\gamma$-ray data and $r$ is the distance from the GC. 

\begin{figure}
    \centering
    \includegraphics[width=0.6\textwidth]{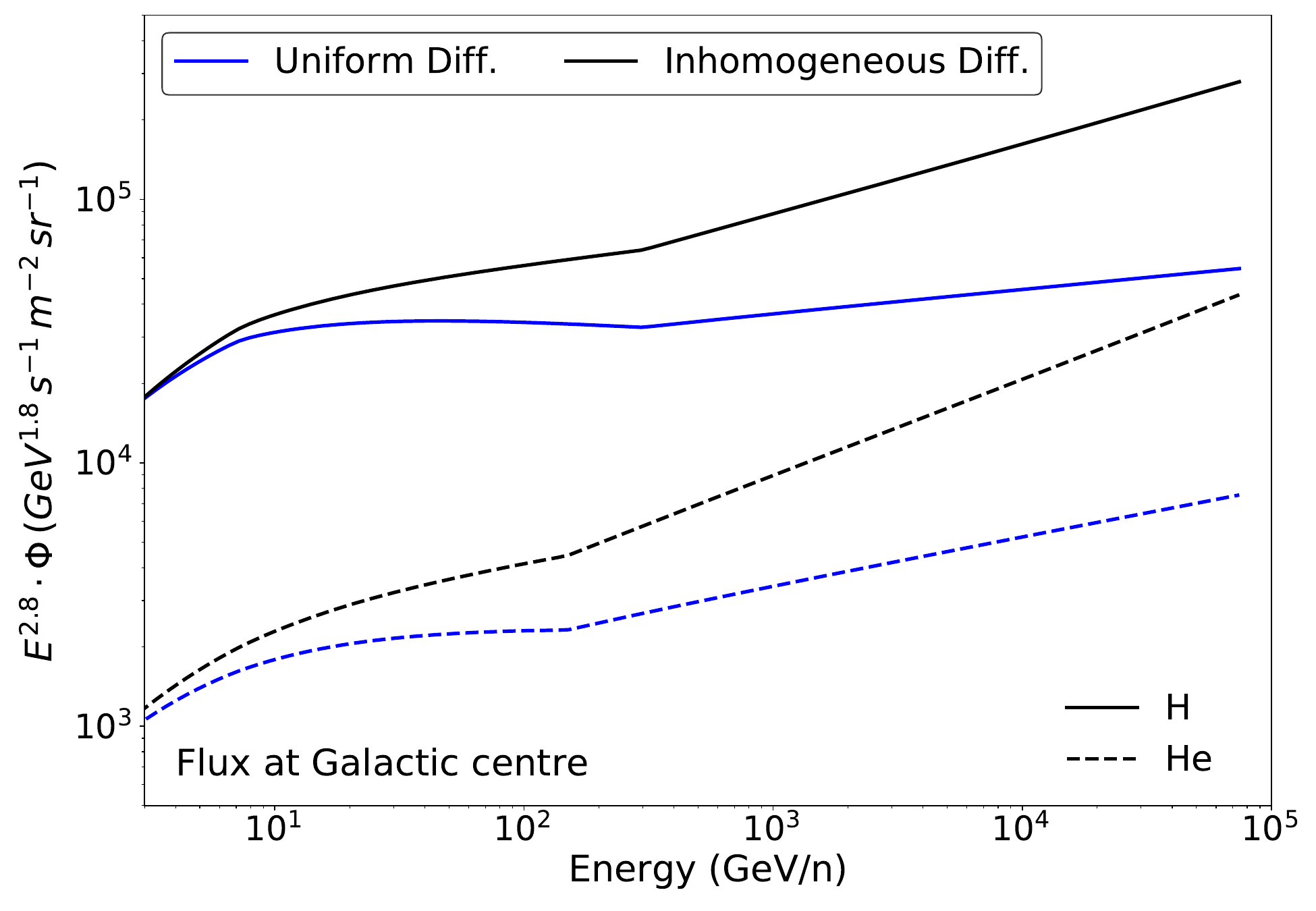}
    \caption{Comparison of the predicted flux of H and He for the inhomogeneous and uniform diffusion models evaluated at the GC.}
    \label{fig:profiles}
\end{figure}

%These non-uniform diffusive models in transport CRs intrinsically implement the density inhomogeneities of our Galaxy. As one approaches the GC, the radial spectral index decreases in value, causing a lower diffusion of the CRs according to equation \ref{eq:1}. 
Physically, the adoption of this non-uniform expression for the diffusion coefficient leads to a higher CR confinement towards the GC (i.e., CRs stay in those zones for longer time, resulting in higher probability of interactions with the ISM and, therefore, more production of secondary particles). In this work, we adopt this parameterization of the spatial dependence of the diffusion coefficient and assess its implications in the current models describing local secondary CR data. Remarkably, the predicted CR flux close to the GC will be very different when assuming uniform and inhomogeneous CR diffusion (while it should not be very different at Earth since we adjust our models to reproduce the local CR data), as we show in Fig.~\ref{fig:profiles} for H and He. Since the flux of local CR particles reflect their transport in the whole Galaxy, we expect secondary particles to show features of inhomogeneous transport.
%Primary nuclei production would also be affected by an inhomogeneous diffusion model. 
%Since the flux of primary CRs follows a power law of the form $\phi_P \sim E^{(\gamma + \delta)}$, where $\gamma$ is the injection parameter (see Eq. \ref{eq:injection}), primary and secondary CRs will show .
%Hence, the spectrum of primary particles would suffer a hardening towards the GC \textcolor{red}{(Ref to Lipari18? Gamma-rays may be associated to this explanation...)}. This phenomenom is illustrated in Figure \ref{fig:profiles}, where H and He nuclei spectra are compared for both homogeneous and inhomogeneous models. Thus, the consideration of an inhomogeneous transport model is of great interest due to the flux discrepancies obtained, supported by gamma ray \textcolor{red}{(Ackermann12, Lipari18, Luque22, etc.)} and, lately, newest neutrino observations \textcolor{red}{(ref IceCube-DiffuseGal). Rephrase this or introduce new sentence. Maybe ask Pedro}.

%This purpose is motivated by the chance of finding substantial changes between both diffusion scenarios that, in the first place, reproduce AMS-02 data with higher accuracy, allowing us to lean in favor of an updated diffusion model. 
While the ratios of secondary-to-primary CR fluxes (such as B/C, Be/O, etc), that are used to determine the propagation parameters, are expected to show small deviations (larger as energy increases) with the inhomogeneous model employed here, the antiproton flux and the antiproton-to-proton ratio (similarly for antinuclei, such as antideuteron and antihelium) could be expected to show larger differences, due to the fact that their flux at a given energy is mainly produced from (around $6$ times) higher energy protons, where the effect of the spatially-dependent propagation setup is more relevant.

For the same reasons, it is worth exploring the scenario of inhomogeneous propagation in the context of the signals expected at Earth from DM production of antiprotons and antinuclei. Considering that most of the DM is concentrated at the GC, one may expect the DM-induced antiproton spectrum to be different in this diffusive scenario.
Yet, in order to reveal anomalies in the antiproton data that could indicate the presence of DM signals, it will be mandatory to pay particular attention to the uncertainties in our predictions because of the diffusion setup employed, which is commonly the uniform diffusion scenario.

%This paper analyzes the impact that a non-homogeneous model of the form of equation \ref{eq:delta} has on the production of secondary particles. In particular, antinuclei $\bar{p}$, $\bar{d}$, $\bar{^3 He}$ production is simulated and compared to those generated in the homogeneous scenario. These antinuclei may be produced as secondary nuclei by spallation or as a result of DM annihilation. Both forms of antinuclei generation will be considered separately for each diffusive model. \pdl{I'd maybe remove this paragraph since we should say this already at the end of the intro}

\label{sec:method}
\subsection{Simulation setup}
\label{sec:simulation}
To simulate the production and arrival of CRs, we solve the full transport equation (Eq.~\ref{eq:transport}), for each CR species $i$, with the \texttt{DRAGON2}\footnote{\texttt{\hyperlink{https://github.com/cosmicrays/DRAGON2\-Beta\_version}{https://github.com/cosmicrays/DRAGON2\-Beta\_version}}} code. This code solves numerically the transport equation within a realistic Galactic environment, yielding solutions of high reliability. For more details about the code, theory and treatment of the transport equation~\citep{DRAGON2_1, DRAGON2_2}.
\begin{equation}
\begin{split}
    -\vec{\nabla} \cdot \left(D\vec{\nabla}N_i+\vec{v}_\omega N_i\right) + \frac{\partial}{\partial p}\left [ p^2 D_{pp}\frac{\partial}{\partial p}\left(\frac{N_i}{p^2}\right)\right] - \frac{\partial}{\partial p}\left[\dot{p}N_i - \frac{p}{3}\left(\vec{\nabla} \cdot \vec{v}_\omega\right) N_i \right] =\\
    = Q + \sum_{i<j} \left(c\beta n_{gas}\sigma_{j\rightarrow i} + \frac{1}{\gamma\tau_{j\rightarrow i}}\right) N_j - \left(c\beta n_{gas}\sigma_{i} + \frac{1}{\gamma\tau_{i}}\right) N_i 
    \label{eq:transport}
\end{split}
\end{equation}

This equation describes the propagation of CRs in the Galaxy for a given species $i$ with density $N_i(\vec{r},p)$. The left-hand-side of the equation describes the diffusive phenomenology (enclosed in the spatial diffusion coefficient, $D$, and $D_{pp}$ in momentum space), as well as the convective processes and energy losses along the path of a CR. The right-hand-side takes into account the sources of injection of CRs for the species $i$, being $Q$ the differential energy flux of primary CRs in units of m$^{-2}$s$^{-1}$sr$^{-1}$GV$^{-1}$, followed by nuclear fragmentation by interaction with the ISM gas (with $Z_j > Z_i$) and lately plausible annihilations of species $i$ or decay into other states.

%The transport equation cannot be solved analytically unless strongly simplified scenarios are assumed, which does not provide realistic solutions. To simulate the production and arrival of CRs, the \texttt{DRAGON2}\footnote{\texttt{\hyperlink{https://github.com/cosmicrays/DRAGON2\-Beta\_version}{https://github.com/cosmicrays/DRAGON2\-Beta\_version}}} code has been used, which solves this equation in a highly realistic galactic environment, providing reliable and accurate solutions. For more details about the code, theory and treatment of the transport equation, see \textcolor{red}{(refs DRAGON2)}.

For the simulations performed in this study, the Galaxy is modeled as a cylinder with azimuthal symmetry and a radius of 30 kpc. Regarding the effective halo size (H, the effective height of the Galactic region where CRs diffuse), we set it to the value obtained in Ref.~\cite{Luque_2021} from analysis of beryllium isotopes ($^{10}$Be), i.e., H=4.72~kpc. In our simulations, we inject the primary CRs $^1$H, $^4$He, $^{12}$C, $^{14}$N, $^{16}$O, $^{20}$Ne, $^{24}$Mg, $^{28}$Si, whose injection is adjusted to reproduce AMS-02 measurements~\cite{AGUILAR20211, Aguilar:2018keu, aguilar2017observation, aguilar2018observation}. %The impact of heavier primary nuclei in the production of secondaries will not affect significantly secondary production and can be neglected, as we will discuss below.
This source injection is parameterized as a doubly broken power-law:
   \begin{equation}
     \label{eq:injection}
     Q_{source} = \left\{
	       \begin{array}{ll}
		 k_1 \times \left(\frac{E}{E_0}\right)^{\gamma_1}      & \mathrm{if\ } E<E_1 \\
		k_2 \times \left(\frac{E}{E_0}\right)^{\gamma_2}      & \mathrm{if\ } E\in [E_1,E_2] \\
		 k_3 \times \left(\frac{E}{E_0}\right)^{\gamma_3}      & \mathrm{if\ } E>E_2
	       \end{array}
	     \right.
   \end{equation}
where $\gamma_{1,2,3}$ are the injection parameters, $k_{1,2,3}$ normalization constants of the flux, and $E_{1,2}$ the rigidity breaks, with $E_1$ = 8.0 GV. $E_2 = 300$~GV. The spatial distribution of sources in the Galaxy is taken from Ref.~\cite{Lorimer_2006}.

The arrival of CRs in the solar environment is subject to the phenomenological conditions of the heliosphere. The Sun follows periods of cyclic activity that vary with time, mainly affecting the flux of low-energy CRs ($\lesssim$ 50 GeV/n) passing through the heliosphere~\cite{Doetinchem_2020,Labrador_1997}. 
%In fact, the flux of low-energy antiprotons on Earth can vary by a factor of two between solar activity maxima and minima \cite{Doetinchem_2020}, being specially affected $\lesssim$ 3 GeV antiprotons \cite{Labrador_1997}.
Solar modulation is parameterized through a modification of the Force-Field approximation~\cite{Potgieter_2013, Gleeson_1967}, derived in Ref.~\cite{Cholis:2015gna}, 
\begin{equation}
    \phi^\pm(t,R) = \phi_0(t) + \phi_1^\pm(t)F(R/R_0),
\end{equation}
 where we set the values of $\phi_0 = 0.61$~GV, $\phi_1$ = $0.90$~GV and $R_0$ to $1$~GV, based on analyses from Ref.~\cite{Luque_2021} and~\cite{Reinert:2017aga}.
 
We will take, as benchmark scenario for the homogeneous model, the diffusion parameters obtained in Ref.~\cite{Luque_2021}, from fits to the AMS-02 fluxes of the primary CRs mentioned above, the ratios of secondaries to primaries (B, Be, Li to C, O) and the antiproton-over-proton ratio, as well as the $^{10}$Be/Be and $^{10}$Be/$^{9}$Be ratios from various experiments -- ACE~\cite{ACEBe}, IMP~\cite{IMP1, IMP2}, ISEE~\cite{ISEE}, ISOMAX~\cite{Hams_2004}, Ulysses~\cite{Ulysses} and Voyager~\cite{VoyagerMO} missions -- (see Ref.~\cite{DelaTorreLuque:2024ozf} for a recent analysis finding similar results). Our setup neglects the high energy break in the diffusion coefficient that was included in Ref.~\cite{Luque_2021} in order to avoid more freedom and degenerate parameters in our evaluations. Instead, we include a break at high energies in the spectra of primary CRs (Eq.~\ref{eq:injection}). We have verified that including this break in the diffusion coefficient only changes the predicted antiproton-over-proton spectrum by a roughly constant normalization factor of $1-2\%$, and the spectrum of B, Be and Li only above $\sim150$~GeV. In this way, the diffusion parameters employed for the (benchmark) homogeneous diffusive model are $\delta = 0.49$,  $D_0 = 4.49\times$10$^{28}$ cm$^2$s$^{-1}$, $\eta = -0.91$ and Alfvén speed of $V_A=13.06$~km/s. These analyses also consider, as nuisance parameters, scaling factors to adjust the normalization of spallation cross sections of secondary nuclei, whose values are compatible with the uncertainties that currently exist~\cite{Genolini_2018}. In particular, the scale factors in our benchmark model, for B, Be, Li and $\bar{p}$ are, respectively, $\mathcal{S}_{B}$ = 0.99, $\mathcal{S}_{Be}$~=~0.92, $\mathcal{S}_{Li}$~=~0.88, $\mathcal{S}_{\bar{p}}$~=~1.11, similar to what is found by the recent analyses of Ref.~\cite{dimauro2023datadriven}. The values of main parameters employed are provided in Table~\ref{table:1}, for each of the propagation setups described in the text.

Then, we consider the inhomogeneous model, evaluated from the diffuse $\gamma$-ray observations by Fermi-LAT, as discussed above, henceforth referred to as \textit{$\gamma$-inhomogeneous}, for comparison with the predictions from the uniform diffusion scenario. The $\gamma$-inhomogeneous model adopts the parameterization of the spectral index found from $\gamma$-ray observations by Ref.~\cite{Luque_PeV}, which has the following form:
\begin{equation}
    \delta(r) = 0.04\cdot r + 0.17
\end{equation}
In this case, we keep the same value for the halo height, $V_A$, $\eta$ and scaling factors, as in the benchmark (homogeneous) scenario, and adjust only the normalization of the diffusion coefficient to fit the B/C ratio from AMS-02 data. The injection spectra of primary CRs is also reevaluated (to reproduce AMS-02 observations) to avoid biasing the results. This allows us to directly compare the impact of such inhomogeneous diffusion setup in the production of secondary CRs (and, specially, in the local antiproton spectrum).
The main propagation parameters employed are summarized in Table~\ref{table:1}, which contains a third model that is explained below (Sec.~\ref{sec:CombSec}).
%It is found that $\gamma$-inhomogeneous can be improved by fitting secondary nuclei spectra. This is because the study \cite{Luque_PeV} shows how the spectral index varies with respect to the galactocentric coordinate, $d\delta(r)/dr = \delta_0$ (see Eq. \ref{eq:delta}), and $\delta_1$ value is inferred by knowing the spectral index value on Earth through the homogeneous model. 

%Regardless of the diffusive model used, the main observables considered are the proton, helium and B/C spectra. Even though here we are considering B/C as the main observable to fit the diffusion parameters, not all boron is produced by carbon. 
%The simulation considers primary particles from $^1$H up to $^{28}$Si. All these injected nuclei produce boron as secondary, which has been taken into account and is implicitly included in the total boron flux used for the B/C ratio. Same applies for other secondaries, such as beryllium or lithium. In parallel, disregarding elements heavier than silicon has an almost negligible impact on secondary particle production. The abundance of heavier elements than Si decreases with the atomic number in the ISM \textcolor{red}{(ref. to the typical plot)}, except for Iron. The production of secondary particles for these heavier elements have an upper limit of 4$\%$ \cite{Genolini_2018}, being no more than 3$\%$ for $\bar{p}$ production \textcolor{red}{(ref?)}. Therefore, considering primary nuclei up to Silicon is sufficient to establish accurate enough predictions of secondary particle production.

We particularly focus our attention to the B/C and $\bar{p}/p$ flux ratios. 
Unlike boron, that forms from spallation interactions of heavy CRs, antiprotons are produced by inelastic collisions of primary particles, being p-p reactions the dominant production channel with $\sim60\%$ contribution to the $\bar{p}$ spectrum~\cite{Luque_PhD}. An important kinematical consequence of the different mechanisms of creation of these secondary CRs is that, while B will roughly inherit the same energy per nucleon as the parent nucleus, antiprotons will be produced at energies much smaller (around a factor of 6) than the parent protons~\cite{Tomassetti_2015}. Therefore, one could expect larger deviations in the predicted $\bar{p}/p$ ratio, from the inhomogeneous scenario, than in the B/C ratio. 
Further, non-annihilating inelastic reactions of the type $p+\bar{p} \rightarrow \bar{p}' + X$ (being $X$ some residual particles) are also allowed, henceforth referred to as tertiary antiprotons, which will be more relevant at low energies. Antinuclei are similarly produced~\cite{DeLaTorreLuque:2024htu}, although in much lower amounts, and we will also discuss how their production can change in the different scenarios. Another source of antiprotons and antinuclei production could be DM annihilation~\cite{DelaTorreLuque:2024ozf, Calore_2022, Genolini_2021}, which is explored in Section~\ref{sec:DM}. 
%We consider WIMPs with a mass from 20 GeV to 1 TeV annihilating through $b\bar{b}$ channel to produce antiprotons, antideuterium and antihelium-3. We compare the spectra obtained in the different scenarios, in order to test if this kind of inhomogeneous propagation could significantly affect our current indirect DM searches.

\renewcommand{\arraystretch}{1.2}
\begin{table}[th!]
\begin{center}
\begin{tabular}{ccc|cccc|}
\multicolumn{1}{l}{}    & \multicolumn{1}{l}{} & \multicolumn{1}{l}{} & \multicolumn{1}{l}{} & \multicolumn{1}{l}{} & \multicolumn{1}{l}{} & \multicolumn{1}{l}{}   \\ \cline{2-7}

\multicolumn{1}{c|}{} & \pmb{$D_0\left[ 10^{28}~cm^2~s^{-1}\right]$} & \pmb{$\delta$} & $\pmb{\mathcal{S}}_B$ & $\pmb{\mathcal{S}}_{Be}$ & $\pmb{\mathcal{S}}_{Li}$ & $\pmb{\mathcal{S}}_{\bar{p}}$  \\ \hline 

\multicolumn{1}{|c|}{\textbf{Homogeneous}} & $4.43$ & \multicolumn{1}{c|}{$0.49$} & \multicolumn{1}{c}{$0.99$} & \multicolumn{1}{c}{$0.96$} & \multicolumn{1}{c}{$0.92$} & \multicolumn{1}{c|}{$1.11$}\\ \hline

\multicolumn{1}{|c|}{\textbf{\pmb{$\gamma$}-inhomogeneous}} &  $4.83$ & \multicolumn{1}{c|}{$0.04\cdot r + 0.17$} & \multicolumn{1}{c}{$0.99$} & \multicolumn{1}{c}{$0.96$} & \multicolumn{1}{c}{$0.92$} & \multicolumn{1}{c|}{$1.11$}\\ \hline

\multicolumn{1}{|c|}{\textbf{Inhomogeneous best-fit}} & $4.55$ & \multicolumn{1}{c|}{$0.04\cdot r + 0.20$} & \multicolumn{1}{c}{$1.00$} & \multicolumn{1}{c}{$0.97$} & \multicolumn{1}{c}{$0.93$} & \multicolumn{1}{c|}{$1.02$}\\ \hline 
\end{tabular}
\end{center}
\caption{Main parameters characterizing the propagation models considered in this work. The normalization and spectral index of the diffusion coefficient, $D_0$ and $\delta$, respectively, are displayed in the central columns, while the scale factors of the cross-section for secondary nuclei are presented in the right columns, $\mathcal{S}_B$, $\mathcal{S}_{Be}$, $\mathcal{S}_{Li}$, $\mathcal{S}_{\bar{p}}$. The rest of propagation parameters are the same in all the scenarios and their values are provided in the main text of Section \ref{sec:method}.}
\label{table:1}
\end{table}

\section{Predictions from the uniform and spatial-dependent propagation setups}
\label{sec:results}

\subsection{Impact on secondary nuclei and antinuclei production}
\label{sec:Sec}

\begin{figure}[t!]
\centering
\includegraphics[width=0.48\linewidth]{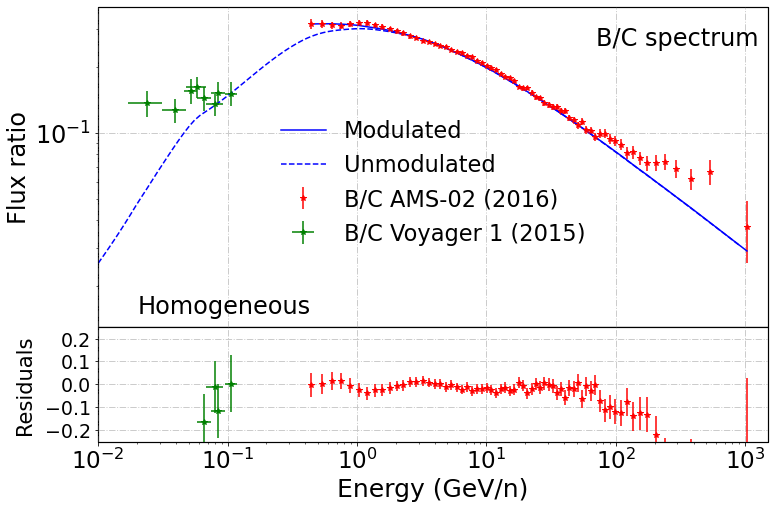} \hspace{0.2cm}
\includegraphics[width=0.48\textwidth]{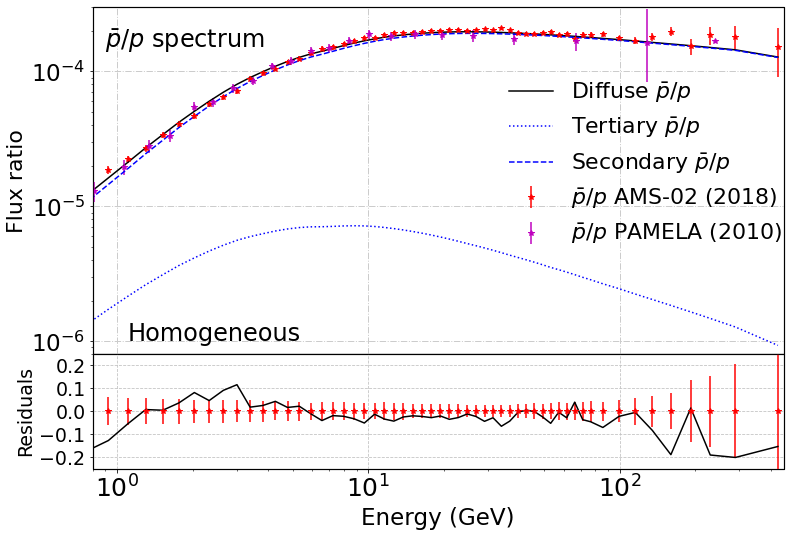}
\includegraphics[width=0.48\linewidth]{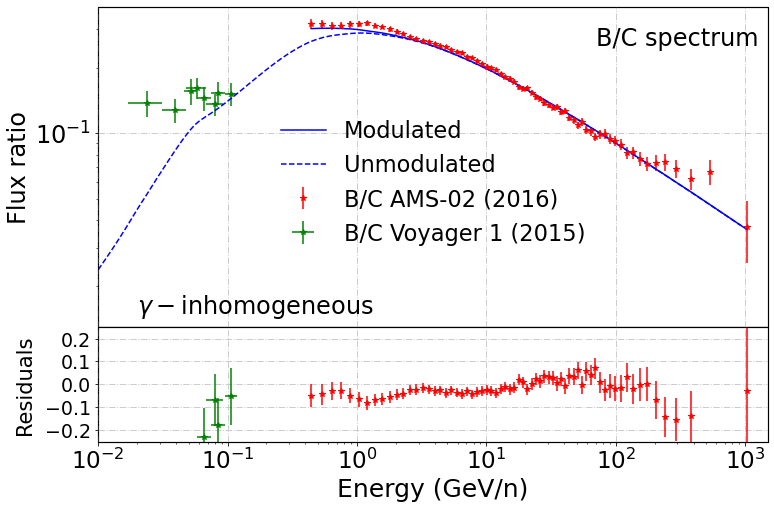}
\hspace{0.2cm}
\includegraphics[width=0.48\textwidth]{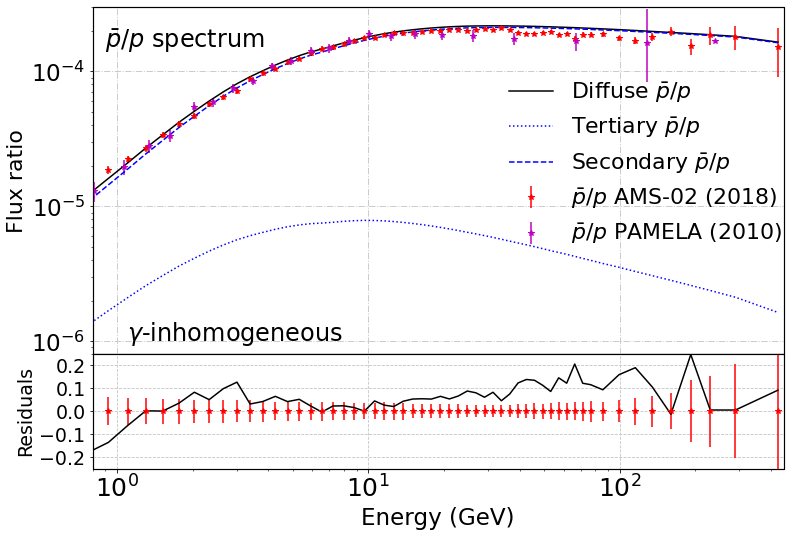}
\caption{\textbf{Left panels:} B/C flux ratios as measured by AMS-02 (red) and Voyager 1 (green), in comparison with our model predictions (blue lines). \textbf{Right panels:} $\bar{p}/p$ flux ratios as measured by AMS-02 (red) and PAMELA (magenta), in comparison with our model predictions (blue lines). The predictions for the diffusive homogeneous model are shown in the top panels and those from the $\gamma$-inhomogeneous in the bottom panels. Residuals are defined as (model$-$data)/data.}
%\caption{\textbf{Upper row:} Predicted B/C (left) and $\bar{p}/p$ flux ratios (right) from the homogeneous diffusion model. \textbf{Lower row:} Predicted B/C and $\bar{p}/p$ flux ratios from the $\gamma$-inhomogeneous model. Residuals are defined as (model-data)/data.}
\label{fig:diff_h}
\end{figure}

In this section, we compare the predicted secondary fluxes (production from CR interactions with the interstellar gas) obtained from the uniform scenario with those obtained from the $\gamma$-inhomogeneous model. In particular, we focus on the B/C and $\bar{p}/p$ ratios. The predicted ratios are compared to AMS-02 data in Fig.~\ref{fig:diff_h} for the homogeneous diffusion scenario (top panels) and $\gamma$-inhomogeneous model (bottom panels). Residuals are shown in the bottom panel, defined as (model$-$data)/data throughout the work, where `model' denotes the predictions obtained via simulation.

Regarding the B/C spectra, we compare the modeled, unmodulated spectrum with Voyager-1 data ~\cite{2013Sci...341..150S} in the MeV/n energy range, and the modulated spectrum with AMS-02 data in the GeV/n-TeV/n range. In the GeV/n range, we observe that the predictions are compatible with AMS-02 data within a $\sim5\%$ in both scenarios, although we observe that above $100$~GeV/n the predicted spectrum in the homogeneous scenario tends to differ more significantly from the data. This has been one of the main arguments motivating the addition of a break in the diffusion coefficient (theoretically explained as a change of turbulence regime~\cite{Blasi_Breaks}) that allows to fit better the secondary-to-primary ratios above a hundred GeV. We remark that in the $\gamma$-inhomogeneous scenario the statistical preference for such a break will be less significant, as can be seen from the residuals.
However, we observe that, although the predicted B/C ratio in the $\gamma$-inhomogeneous scenario appears compatible with the $1\sigma$ uncertainties in AMS-02 data, the predicted trend of the spectrum seems slightly harder than the one observed in the data. 

Regarding the $\bar{p}/p$ spectrum, we observe that the predicted ratio in the homogeneous scenario seems to follow a quite similar trend to that from AMS-02 data, showing very flat residuals. 
Moreover, above around $100$~GeV, the predicted antiproton spectrum seems slightly softer than that from AMS-02 data. This does not improve adding a high-energy break in the diffusion coefficient and has been observed in the past, motivating the proposal of different mechanisms to explain this disagreement in the trends~\cite{Cholis_StocAcc, Mertsch_AccSec}. Note, though, that it is not a statistically significant issue, given the uncertainty in antiproton data at those energies.
In turn, the $\bar{p}/p$ flux ratio obtained from $\gamma$-inhomogeneous model clearly reveals a spectrum that deviates significantly from AMS-02 measurements, with a discrepancy that grows as energy gets higher. In particular, the predicted $\bar{p}/p$ ratio in this scenario shows a higher discrepancy with respect to the trend measured by AMS-02 than in the B/C ratio. This diffusion scenario entails, therefore, a substantial increase in the production of antiprotons.
For the sake of a fair comparison, modulation parameters as well as cross sections scale factors are the same as in the homogeneous scenario. In fact, what we appreciate is not only a larger production of antiprotons, but a remarkable change in the shape of the spectrum, particularly at high energies.
It is important to notice that modifying the scale factors for antiproton cross sections could get our prediction in the $\gamma$-inhomogeneous scenario closer to the data, but yet the trend of the spectrum and the data will still be very different and incompatible with the measurements. 
%It is also found that tertiary production of $\bar{p}$ makes an non-negligible contribution at $\sim$10 GeV. 
%One can observe how the predicted antiproton spectrum below around 3 GeV starts to differ with the trend from AMS-02 data. This is probably related to the impact of Solar modulation and the uncertainties related to their modelling, according to Ref.~\cite{Labrador_1997}.

\begin{figure}[t!]
    \centering
    \includegraphics[width=0.7\textwidth]{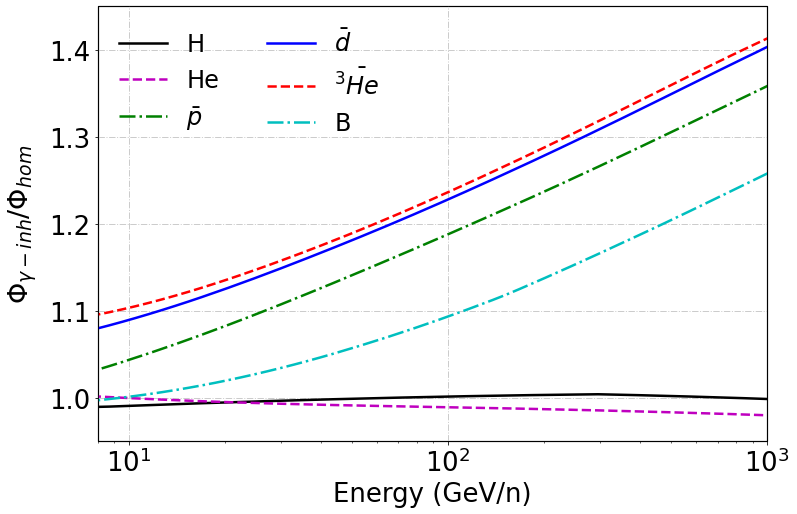}
    \caption{Ratios of the predicted fluxes at Earth between the $\gamma$-inhomogeneous diffusive model and the homogeneous model, for different nuclei depicted in the legend.}
    \label{fig:Ratios}
\end{figure}

To facilitate the comparison of the differences between the considered diffusive models in the production of secondary CRs, we illustrate the ratio with energy between both scenarios in Fig.~\ref{fig:Ratios} up to $1$~TeV/n for protons, He, B, $\bar{p}$, $\bar{d}$ and $^3\bar{He}$ nuclei. %The set ups for both models are those already discussed. 
We find that the production of secondary nuclei increases substantially in the $\gamma$-inhomogeneous model, predicting a harder trend of the spectrum. For instance, the predicted B flux is about $10\%$ greater than in the homogeneous case at $100$~GeV/n. 
Concerning antinuclei, fluxes of $\bar{p}$, $\bar{d}$, $\bar{^3He}$ are around $20\%$ greater than in the homogeneous case at $100$~GeV/n. 
This larger increase of secondary antiprotons and antinuclei with respect to B is due to the fact that the spectrum of antiprotons at this energy is mainly produced by protons of much higher energies, where the differences from the inhomogeneous and homogeneous scenarios become more significant. The heavier the antinuclei, the larger deviation due to the kinematics of the interactions producing them ($\bar{d}$ are mainly produced by protons with energy around $15$ times higher, and $\bar{^3He}$ is mainly formed from protons around $30$ times more energetic). 

These comparisons also indicate that although uncertainties in the determination of the spatial dependence of the diffusion coefficient can be relevant in our predictions of CR spectra at Earth, cross section uncertainties on the production of secondary CRs are still the leading factor of uncertainty\footnote{For CR secondary nuclei and antiprotons, these uncertainties are often evaluated to be around $20\%$, and even higher above tens of GeV, due to the lack of experimental data. For antinuclei, the lack of measurements do not allow us to have a robust estimation and different evaluations agree usually within a factor of $2$ for $\bar{d}$ and at least a factor of a few for $\bar{^3He}$.}, not allowing us to really distinguish between these propagation models with any accuracy.
% We first quantify and estimate the level of deviation in our predictions when including  inhomogeneous diffusion. This model has some uncertainties related, for sure, but they are much smaller than those related to cross sections uncertainties.

\subsubsection{Combined fits in the inhomogeneous setup}
\label{sec:CombSec}
\begin{figure}[t]
    \centering
    \includegraphics[width=0.495\textwidth]{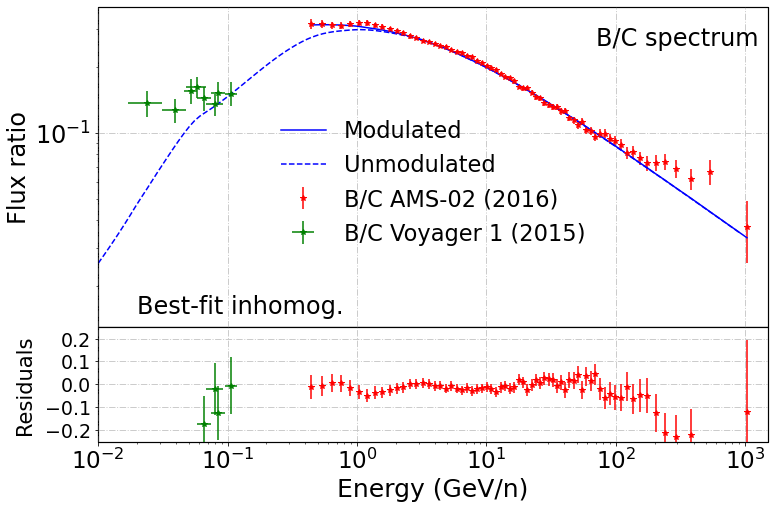}
    \includegraphics[width=0.495\textwidth]{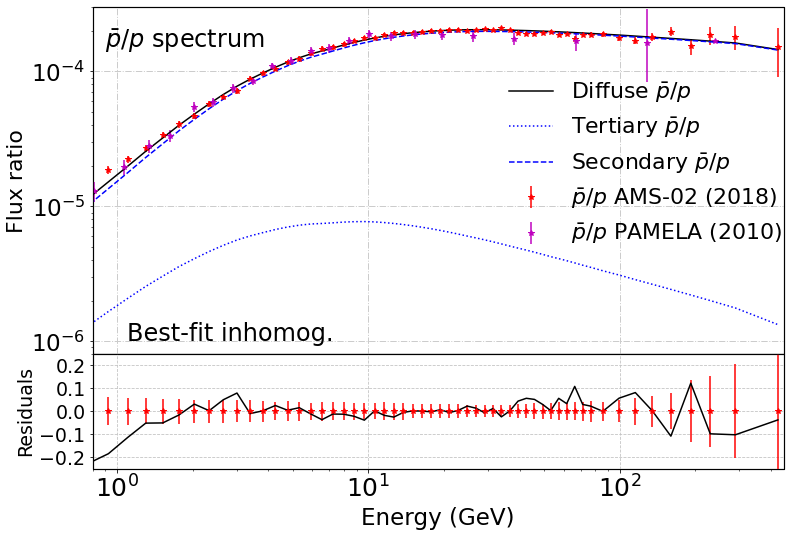}
    \caption{\textbf{Left panel:} B/C flux ratio measured by AMS-02 (red) and Voyager 1 (green), in comparison with the prediction from our Best-fit inhomogeneous diffusion model (blue lines). \textbf{Right panel:} $\bar{p}/p$ flux ratio measured by AMS-02 (red) and PAMELA (magenta) compared with that predicted from best-fit inhomogeneous diffusion model. Residuals are defined as (model$-$data)/data.}
    \label{fig:ImhFit_main}
\end{figure}

Given that the $\gamma$-inhomogeneous model seems to predict spectra whose trend for the B/C and $\bar{p}/p$ ratios do not exactly follow that from AMS-02 data (specially for the $\bar{p}/p$ ratio, the predicted spectrum seems considerably harder; see bottom left panel of Figure \ref{fig:diff_h}), we address here if an overall reproduction of the B, Be, Li and $\bar{p}$ species can be achieved simultaneously with the kind of inhomogeneous model that we are using. 
In particular, we adjust the spectra of these species through the simultaneous fit of the propagation and injection parameters to AMS-02 data, leaving fixed the $\delta_0$ parameter (see Eq.~\ref{eq:delta}), since it is obtained from $\gamma$-ray diffuse observations, as well as the $\eta$ and $V_A$ parameters. Therefore, we let free in the fit the $D_0$ and $\delta_1$ (which regulates the $\delta$ value at Earth position), as well as the scale factors for the cross sections of these secondary CRs. This fit allows us to reproduce the AMS-02 data accurately, as shown in Fig.~\ref{fig:ImhFit_main} for the $\bar{p}/p$ (right panel) and B/C (left panel) ratios, as well as for the rest of CRs (see Fig.~\ref{fig:Spectra} in Appendix~\ref{sec:AppA}).
The obtained best-fit parameters obtained are given in Table~\ref{table:1}. We also show the predicted secondary spectra of $\bar{d}$, $\bar{^3He}$ in Appendix~\ref{sec:AppB}, compared to current upper limits and sensitivities of future detectors.\newline
Two main conclusions can be derived from our analysis of the \textit{best-fit} $\gamma$-inhomogeneous model:
\begin{itemize}
    \item The combined fit in the inhomogeneous scenario leads to a best-fit scaling of the antiproton cross sections of $2\%$, much smaller than the $11\%$ scale needed in the homogeneous case (rightmost column of Table \ref{table:1}). This means that this scenario significantly reduces the tension in the predicted grammage needed to reproduce the antiproton spectrum with respect to the grammage needed to reproduce the group of B, Be and Li.
    \item We observe that the antiproton spectrum (and the $\bar{p}/p$ ratio) gets slightly harder, as observed before. This may alleviate the fact, also commented above, that the high energy part of the predicted antiproton spectrum seems softer than that observed by AMS-02, making again more compatible the predictions from B, Be, Li and $\bar{p}$ in the scenario where only secondary production is assumed.
\end{itemize}

Concretely, we observe that the predicted local B spectrum only increases by $\sim3\%$ at $100$~GeV with respect to the benchmark homogeneous diffusion setup, while the local $\bar{p}$ spectrum changes up to a $7-8\%$ at this energy (see Fig.~\ref{fig:Ratios_BF} in App.~\ref{sec:AppA}). Given that the uncertainties in the current data above $100$~GeV are still high, it is not possible to find clear indications favoring any scenario and it would be too premature to state that this kind of inhomogeneous model makes our predicted antiproton spectrum from secondary production significantly more compatible with that of B, Be and Li compared to predictions from a homogeneous model. A recent work~\cite{dimauro2023datadriven} reached a similar conclusion when analyzing other kinds of inhomogeneous setups. We do observe that this kind of inhomogeneous scenario, motivated by diffuse $\gamma$-ray observations, can offer a simple and realistic explanation for the tensions discussed above, and that future data on secondary CRs at higher energies could lead to more robust indications favoring a concrete propagation scenario. In general, we observe that different kind of inmohogeneous models proposed in the literature (see e.g. Refs.\cite{Tomassetti_2012, recchia2023origin, Guo_2016}) lead to a predicted $\bar{p}$ spectrum more compatible with AMS-02 data than homogeneous models. This could indicate a common trait: models that predict a proton spectrum slightly harder than locally outside our local neighborhood are favored by the current $\bar{p}$ local measurements.

\subsection{Impact on possible dark matter signals at Earth}
\label{sec:DM}

In this section we explore the impact that the different considered diffusion setups may have in the production of antiprotons from the annihilation of DM particles. The DM particle, $\chi$, can annihilate producing $b\bar{b}$ quarks which eventually decay into other hadrons, producing antiprotons as a final result ($\chi + \chi \rightarrow b\bar{b} \rightarrow \bar{p} + X$). Even though other annihilation channels are kinematically available~\cite{Marco_Cirelli_2011, arina2023cosmixs}, we take $b\bar{b}$ as the reference channel hereafter. We stress that a different annihilation channel would affect equally the predicted spectra in both scenarios, since it changes only the spectrum at production, leaving unchanged our conclusions.
\begin{figure}[t]
    \centering
    \includegraphics[width=0.495\textwidth]{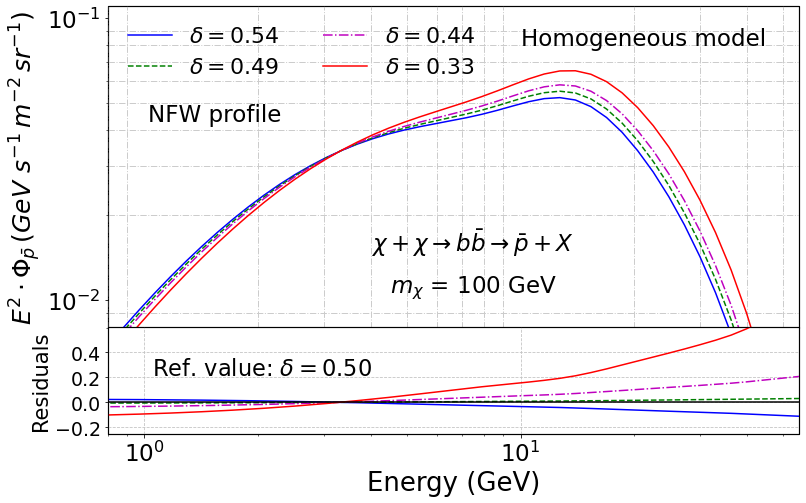}
    \includegraphics[width=0.495\textwidth]{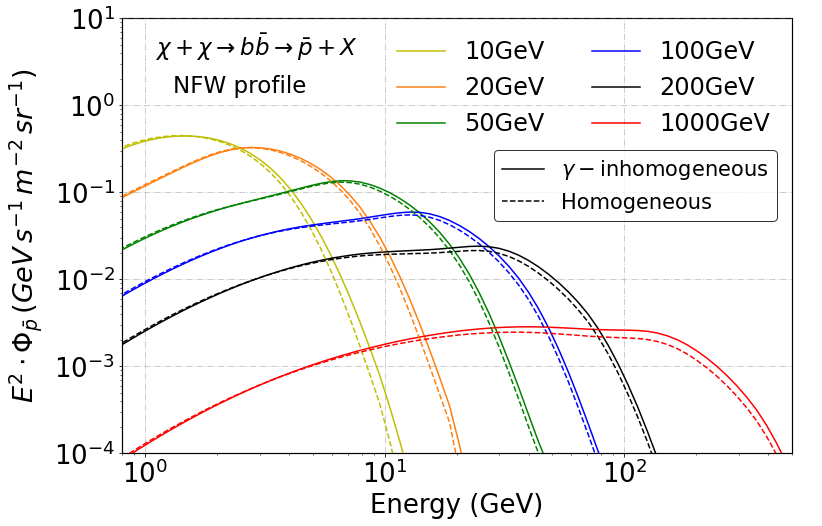}
    \caption{\textbf{Left panel:} Impact on $\bar{p}$ production from DM annihilation for different effective spectral indices in the homogeneous propagation model. Residuals are taken as (model$-$ref.model)/ref.model, where ref.model is the flux obtained with $\delta = 0.5$. \textbf{Right panel:} Comparison of the predicted $\bar{p}$ flux from DM annihilation for various WIMP masses in different propagation scenarios. The solid lines represent the flux predicted by the $\gamma$-inhomogeneous model, while the dashed lines represent those from the homogeneous model. The annihilation rate is set to $\langle \sigma v \rangle$ = 2.33$\cdot$10$^{-26}$ cm$^3$/s, i.e., the thermal relic value~\cite{Steigman_2012}. In both panels we adopt an NFW profile~\cite{Navarro:1995iw} for the DM distribution in the Galaxy.}
    \label{fig:DM_delta}
\end{figure}

As one would expect, the measured antiproton flux on Earth is sensitive to the parameters of the transport model. Different diffusion parameters will therefore provide different DM signals. 
As we have discussed throughout the work, a different spectral index indicates that particles of different energies will diffuse for longer or shorter time in the Galaxy. Similarly, if one considers the $\gamma$-inhomogeneous model (Eq.~\ref{eq:delta}), which implies that particles spend longer times at the GC, it is expected that the effect in the DM signal at Earth would be equivalent to that predicted from a low effective spectral index (i.e., $\delta_{effective}^{inhom.} < \delta^{hom.}$). This reasoning gets even stronger given that the density of DM particles is expected to be higher in the GC.
In the left panel of Figure \ref{fig:DM_delta}, we show the predicted $\bar{p}$ spectrum from DM annihilation in the homogeneous scenario, for different values of the spectral index, all of them plausible and consistent with the CRs transport theory. For instance, a variation of the effective spectral index of $\Delta\delta \approx 0.20$ changes the $\bar{p}$ signal at Earth approximately by a factor of $2$. In this exercise, we adopted the standard Navarro-Frenk-White density profile~\cite{Navarro:1995iw} for the distribution of DM in the Galaxy.

Hence, given the form of the spectral index considered in the inhomogeneous models, we expect a higher local flux of antiprotons and antinuclei from DM than in the homogeneous model. This is corroborated by Figure~\ref{fig:DM_delta} (right), where we perform a comparison of the $\bar{p}$ local flux obtained in the $\gamma$-inhomogeneous model with respect to the homogeneous scenario. WIMP masses of $10$, $20$, $50$, $100$, $200$ and $1000$~GeV are considered to annihilate to $b\bar{b}$ with a cross section equal to the thermal relic cross section, with value $\langle \sigma v \rangle$ = 2.33$\cdot$10$^{-26}$ cm$^3$/s~\cite{Steigman_2012}. A comparison of both models reveals that the local $\bar{p}$ flux can be at most about $25\%$ higher in the $\gamma$-inhomogeneous model at the peak of the emission. The discrepancy between the maximum flux produced by both models becomes more pronounced as the WIMP mass increases.
In the range of WIMP masses between $10$ and $100$~GeV, the inhomogeneous models predict fluxes that are higher than those predicted in the homogeneous scenario by $1-10\%$. %For higher masses, differences of 15$\%$ (m$_\chi$ = 200 GeV), 25$\%$ (m$_\chi$ = 1000 GeV) are observed at these spectral regions. \dlt{For regions outside the spectral peak, we find a $10-15\%$ increase in $\bar{p}/p$ predicted flux ratio. Therefore, with the $\gamma$-inhomogeneous model an increase in antiprotons produced by DM annihilation ranging from $10$ to $20\%$ is found.} 
In addition, one would expect that these differences grow if we consider a more peaked DM density profile (more DM in the inner Galaxy), such as the contracted NFW (c-NFW), with slope of $\gamma=1.2$, proposed to fit the Galactic center excess observed in $\gamma$-rays, see e.g.,~\cite{Ackermann_2017, Calore_2015}. %Also if a large amount of DM is concentrated around SgrA*, we would expect to have such kind of effect.

\begin{figure}[t]
    \centering
    \includegraphics[width=0.49\textwidth]{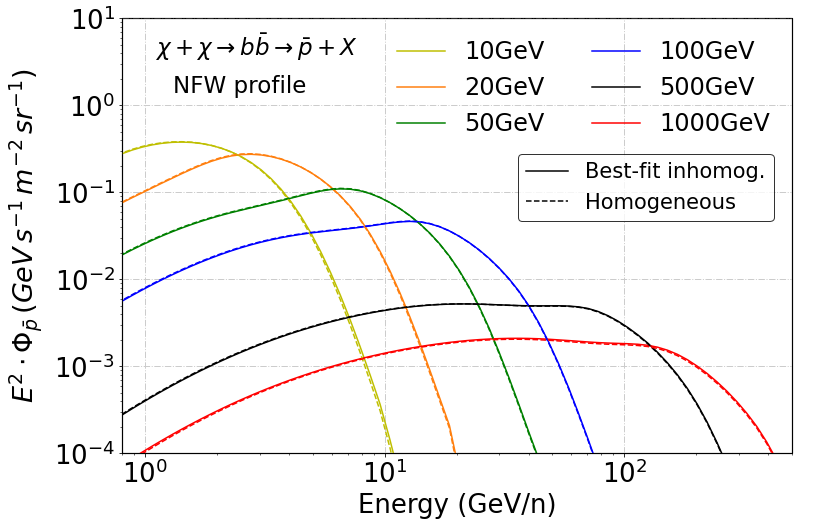}
    \includegraphics[width=0.49\textwidth]{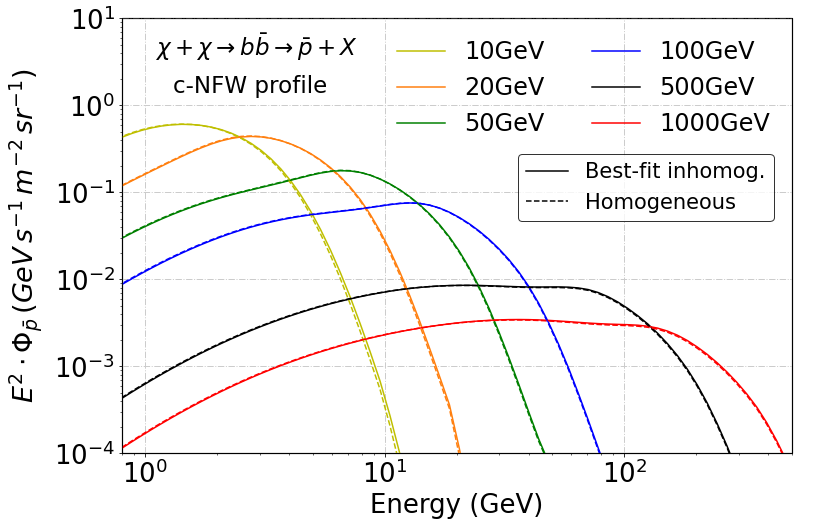}
    \caption{Comparison of the flux of antiprotons produced from DM annihilation for the case of adopting a NFW (left panel) and a c-NFW (right panel) DM density profile for the Galaxy, evaluated in the setup of uniform diffusion and with the best-fit setup of inhomogeneous diffusion.}
    \label{fig:DMAp}
\end{figure}

We also evaluated the DM-induced local flux of antiprotons with the best-fit inhomogeneous scenario. What we found is that no important difference is found between the flux predicted in the homogeneous and best-fit inhomogeneous cases, for any of the DM masses explored. We illustrate this in Figure~\ref{fig:DMAp}, where we consider two different DM density profiles: the NFW (left panel) and the c-NFW (right panel). In either case, the discrepancy in the antiproton flux between both models does not exceed $5\%$. %In the spectral peaks, we observe differences that barely reach 3$\%$, both originating from the c-NFW profile for masses $m_\chi$ = 500, 1000 GeV. No more than 1-2$\%$ is found in the spectral peaks of the NFW profile. The discrepancies between both diffusive models slightly increase in the spectral tails, where the antiproton flux drops dramatically, without exceeding 5-6$\%$.
This demonstrates that the constraints provided by local (secondary) CR data allow us to have a robust estimation of the DM-induced signals at Earth, regardless of how diffusion is in other places of the Galaxy. Yet, a larger family of inhomogeneous diffusion models must be explored to confidently establish this in every case.

\section{Discussion and conclusions}
\label{sec:conc}
In this study, we have examined the implications of assuming an inhomogeneous diffusion model of CR transport in the local flux of different secondary CRs in the GeV/n-TeV/n energy range, focusing on antiproton ($\bar{p}$) production and secondary nuclei. Given that the main information that we have from CRs is their local flux, the scientific community has predominantly explored uniform diffusion models to understand their propagation process. The main motivation to study an inhomogeneous transport model stems from recent results using neutrinos~\cite{IceCube} and $\gamma$-rays~\cite{Cao_2023, TIBET, Fermi_Gamma}, which have been shown to be successfully explained through the adoption of a spatially-dependent diffusion coefficient~\cite{Gaggero_2015, Luque_PeV}. 
In this paper, we particularly focus on the differences between the predicted local flux of secondary CRs, antiprotons and antinuclei as given by homogeneous diffusion scenarios, and a class of inhomogeneous propagation model that considers a galactocentric radial dependence of the diffusion coefficient. This is investigated for both, their production in CR interactions with the ISM gas, and as products of DM annihilation (the latter only in the case of antiprotons and light antinuclei).

A general trait of inhomogeneous propagation models is that the slope of the spectrum of secondary CRs smoothly varies with energy (i.e., the spectrum is not a simple power-law), because different energies feature the propagation of CRs in different effective zones of the Galaxy (see, e.g., Ref.~\cite{Tomassetti_2012}). However, these changes may be arbitrarily small, as the inhomogeneous model explored here predicts. Therefore, we have conducted a direct comparison of the predictions from the homogeneous and inhomogeneous scenarios to understand what are the features that may lead us to disentangle between these scenarios. To achieve our goal, we have solved the full transport equation within a realistic Galactic, state-of-the-art environment using the \texttt{DRAGON2} code. We first make sure to reproduce AMS-02, PAMELA and Voyager-1 experimental data of primary and secondary nuclei, and secondary-to-primary flux ratio, with simultaneous fits in each scenario to guarantee highly reliable and consistent models.

Regarding the secondary production of CRs, we find that the inhomogeneous model predicts a remarkably harder spectrum at Earth compared to the homogeneous propagation case, yet only marginally different to the one predicted by the homogeneous model in the GeV/n range once we adjust the model to reproduce simultaneously all the AMS-02 measurements (specially, the observations on secondary B, Be, Li and $\bar{p}$). However, we demonstrate that the production of secondary antiprotons is significantly more affected by inhomogeneous propagation (due to kinematical effects of the interactions producing them), leading to higher production at higher energies. In particular, the antiproton TeV flux predicted in this scenario would become significantly different to that expected in the homogeneous scenario. Therefore, should future data confirm the current hint -- i.e., that the antiproton data seems to follow a harder trend than the one predicted in current (homogeneous) models -- this could serve as an indication in favor of inhomogeneous propagation of CRs in the Galaxy. 
An additional point favoring the inhomogeneous scenario is that we are able to achieve a very good simultaneous fit of B, Be, Li and $\bar{p}$ almost without modifying the antiproton cross sections, while this simultaneous fit in the homogeneous scenario requires around a $10\%$ scaling of the antiproton cross sections to reproduce all these species simultaneously.

We also investigate how this inhomogeneous diffusion model affects the predicted local flux from DM production of antiproton. We find that, although one could expect small variations on the predicted flux of these antiparticles at Earth, the current observations of secondary CRs by AMS-02 allow us to constrain these DM signals at Earth to be roughly the same in both scenarios (always lower than $\sim5\%$ difference). We find that this is true even for a very peaked DM density profile for the Galaxy, such as the contracted NFW profile that fits the $\gamma$-ray Galactic center excess, and for a large range of DM masses that we test here (from $\sim10$~GeV to $\sim 1$~TeV). Yet, we expect our current DM indirect searches with antiprotons must be affected, at least, by the expected secondary $\bar{p}$ production. Although our results are based only in the b$\bar{b}$ channel, the same conclusions must hold for other channels too, given that changing to a different channel only affects the CR spectrum at the injection.

To end, we emphasize that the propagation of CRs in the Galaxy can be more complex than initially thought, and that different kind of inhomogeneous (and anisotropic) diffusion scenarios will result in different key features that could be tested with current and future CR experiments. Indeed, probably only a combination of different observables and messengers would allow us to improve our current description of the CR propagation processes in the Galaxy in the near future.

%\newpage
\acknowledgments
We thank Daniele Gaggero for useful discussions. P.D.L. has been supported by the European Research Council under grant 742104 and the Swedish National Space Agency under contract 117/19 and is currently supported by the Juan de la Cierva JDC2022-048916-I grant, funded by MCIU/AEI/10.13039/501100011033 European Union "NextGenerationEU"/PRTR. The work of P.D.L. and M.A.S.C. was supported by the grants PID2021-125331NB-I00 and CEX2020-001007-S, both funded by MCIN/AEI/10.13039/501100011033 and by ``ERDF A way of making Europe''. They also acknowledge the MultiDark Network, ref. RED2022-134411-T.
This project used computing resources from the Swedish National Infrastructure for Computing (SNIC) projects 2021/3-42, 2021/6-326, 2021-1-24 and 2022/3-27 partially funded by the Swedish Research Council through grant no. 2018-05973.

%\clearpage
\appendix

\section{Other observables in the best-fit inhomogeneous model}
\label{sec:AppA}

\begin{figure}[h!]
    \centering
    \includegraphics[width=0.9\textwidth]{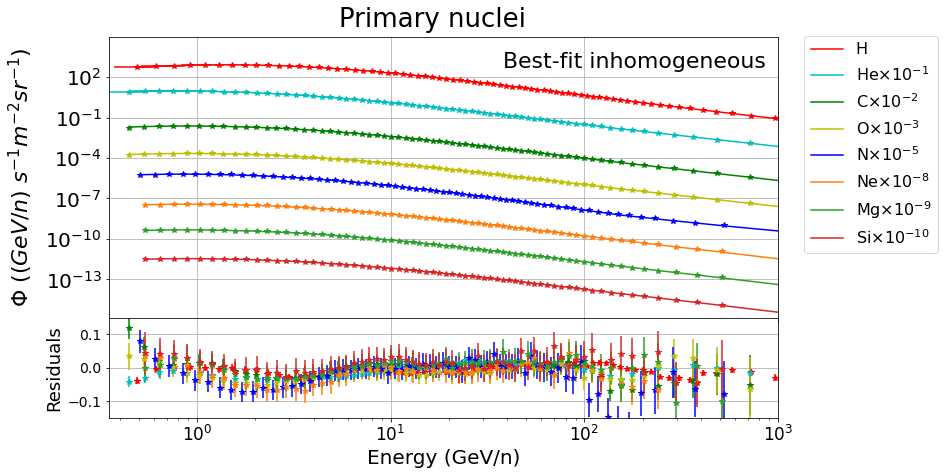} 
    \includegraphics[width=0.9\textwidth]{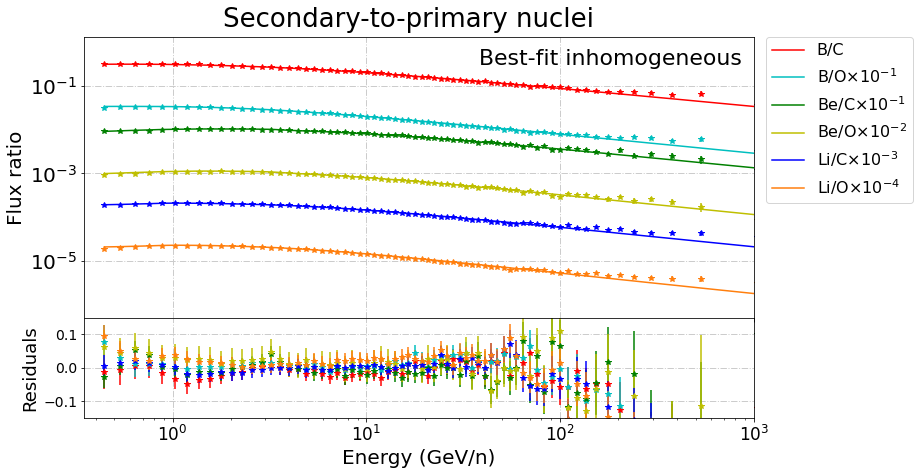}
    \caption{\textbf{Upper panel}: Primary nuclei spectra at Earth measured by AMS-02 (points) and predicted by the best-fit inhomogeneous model (solid line). \textbf{Lower panel:} Secondary-to-primary nuclei flux ratio at Earth measured by AMS-02 (points) and predicted by the best-fit inhomogeneous model (solid line). Residuals are defined as (model - data)/data.}
    \label{fig:Spectra}
\end{figure}

We show in this appendix the fitted spectra of primary and secondary CRs in the best-fit inhomogeneous model, depicted in Table \ref{table:1} and Section \ref{sec:CombSec}. 
In Figure \ref{fig:Spectra}, we illustrate the simulated spectra of p, He, C, N, O, Ne, Mg and Si (top panel) and the ratios of B, Be and Li to C and O (bottom panel), which are fitted to AMS-02 and PAMELA data in the GeV/n energy range. The fit of the primary nuclei provides the injection parameters for each species, according to Eq. \ref{eq:injection}. The fit to the secondary-to-primary flux ratios, along with the $\bar{p}/p$ ratio, allows us to infer the diffusion parameters in agreement with experimental data. %These fits are performed simultaneously and ensure a prediction of antiproton and other secondaries as realistic as possible.

\begin{figure}[h!]
    \centering
    \includegraphics[width=0.75\textwidth]{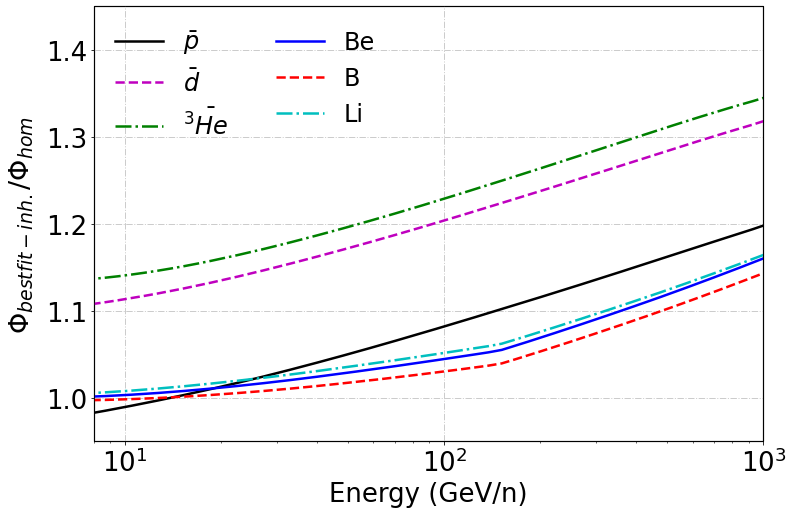}
    \caption{Comparison of the ratios of predicted CR fluxes at Earth from the best-fit inhomogeneous diffusive model and the homogeneous model, for secondary CR nuclei and antinuclei. }
    \label{fig:Ratios_BF}
\end{figure}

In Figure \ref{fig:Ratios_BF}, a comparison of nuclei and antinuclei fluxes is established between the best-fit inhomogeneous model and the homogeneous one. The first one predicts a production of secondary nuclei (B, Be, Li) approximately 15$\%$ higher than the homogeneous model at 1 TeV/n, with an increase of about 20$\%$ for antiprotons and 30-35$\%$ for antideuterium and antihelium-3. Nevertheless, the production of secondaries decreases by approximately 10$\%$ compared to the $\gamma$-inhomogeneous model, with antiprotons flux being the most affected (from 35$\%$ down to 20$\%$).

\section{Antinuclei predictions}
\label{sec:AppB}

\begin{figure}[h!]
    \centering
    \includegraphics[width=0.495\textwidth]{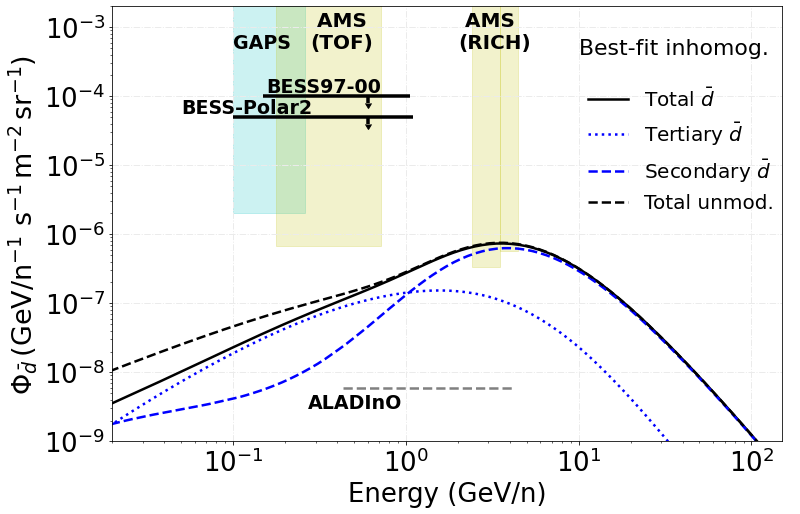}
    \includegraphics[width=0.495\textwidth]{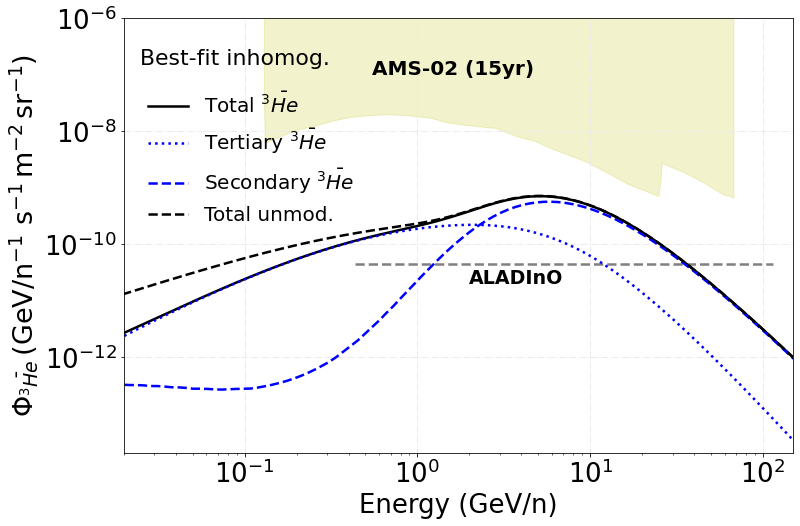}
    \includegraphics[width=0.495\textwidth]{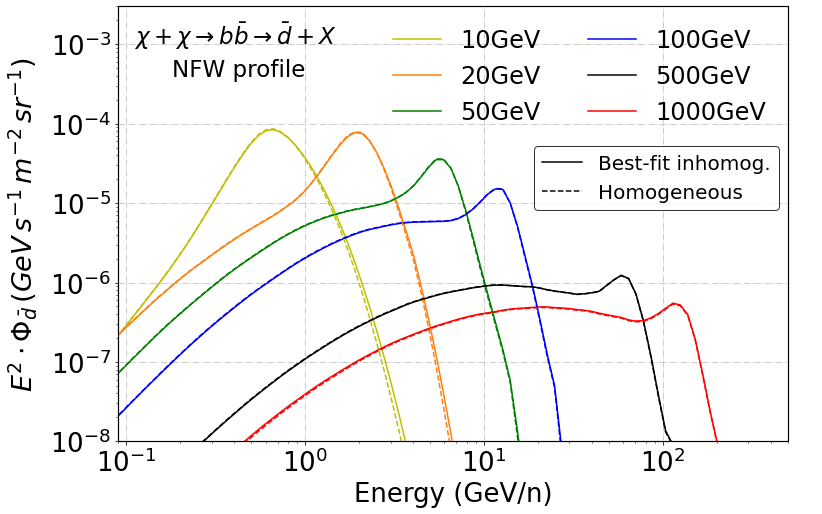}
    \includegraphics[width=0.495\textwidth]{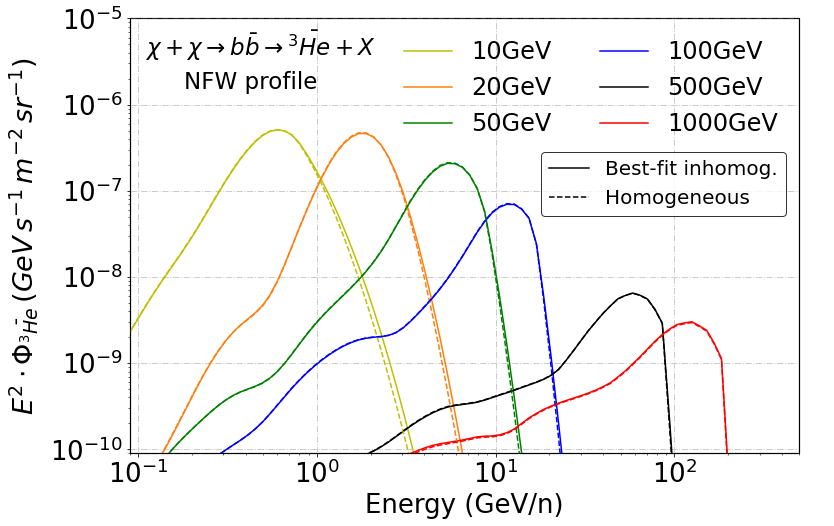}
    \includegraphics[width=0.495\textwidth]{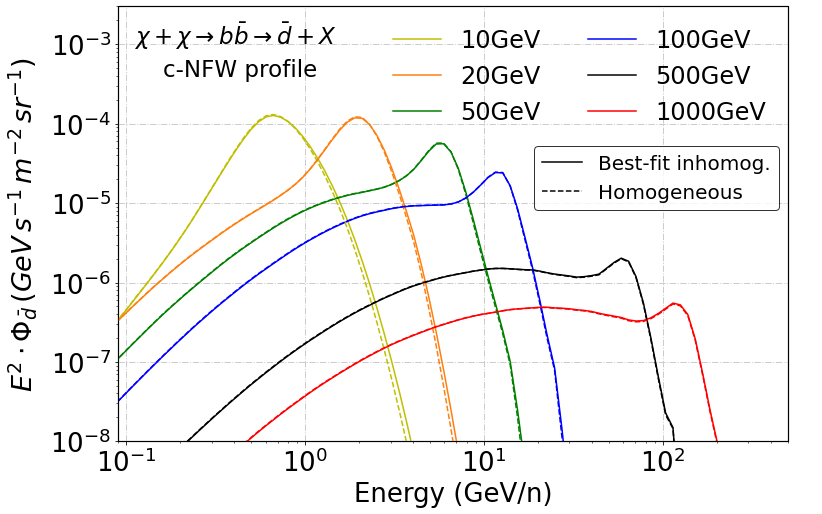}
    \includegraphics[width=0.495\textwidth]{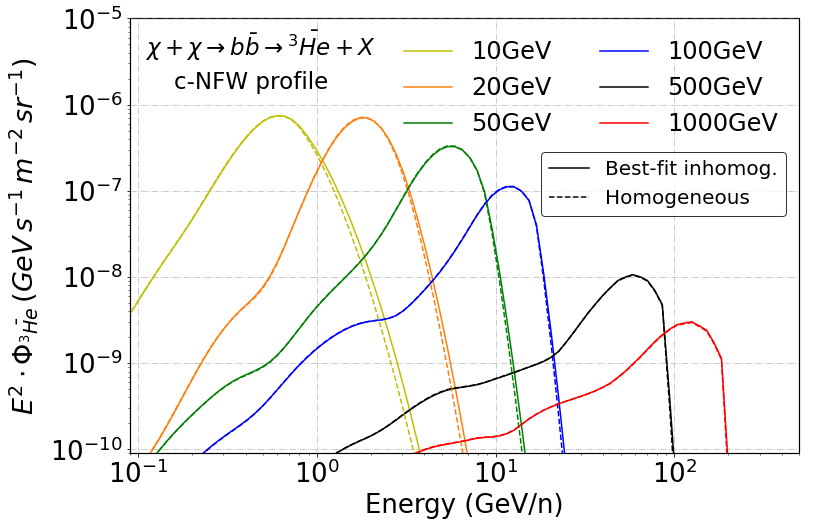}
    \caption{\textbf{Top panels:} Secondary antinuclei $\bar{d}$, $\bar{^3He}$ predicted fluxes with sensitivity bands of future experiments. No $\bar{^3He}$ from CR interactions is expected to be measured in the near future, whereas 1-4 GeV/n $\bar{d}$ enter in the sensitivity domain of AMS (RICH). \textbf{Middle and bottom panels:} $\bar{d}$, $\bar{^3He}$ predicted fluxes within best-fit inhomogeneous model (solid line) from DM annihilation for NFW (central panels) and c-NFW (bottom panels) profiles compared to the fluxes predicted within the homogeneous model (dashed line). No differences higher than 3$\%$ in flux are obtained between both models at the spectral peaks. In most cases, solid lines overlap dashed lines.}
    \label{fig:ImhAN}
\end{figure}

In this appendix, we present and delve further into the details of antinuclei spectra. The full inplementation of antinuclei production and propagation in the DRAGON2 code is described in Refs~\cite{DeLaTorreLuque:2024htu, DelaTorreLuque:2023vvo}.
In the context of the best-fit inhomogeneous model, we illustrate the predicted fluxes for antideuterium ($\bar{d}$) and antihelium-3 ($\bar{^3He}$) in Figure \ref{fig:ImhAN}. Antinuclei can be produced as secondary CRs (top panels) and as the end product of DM annihilation through the $b\bar{b}$ channel for NFW (middle panels) and c-NFW (bottom panels) DM density profiles.
Currently, there are no experimental data available to contrast the predicted fluxes. In turn, we present the sensitivity bands of specific experiments that can detect such antinuclei. In the top panels of Fig. \ref{fig:ImhAN}, we show the sensitivity regions for the RICH and TOF instruments in the AMS-02 detector (taken from Ref.~\cite{ARAMAKI20166}), for $15$~years of operation, the upper-limits from the Balloon-borne Experiment with a Superconducting Spectrometer (BESS)~\cite{BESS_upper}, and the sensitivity region for the General AntiParticle Spectrometer (GAPS)~\cite{ARAMAKI20166, GAPS_Ap} (for the expected three flights of $35$~days, each). Moreover, we include the forecasted sensitivity for the future Antimatter Large Acceptance Detector In Orbit (ALADInO)~\cite{aladino} (expected for $5$~years of operation). According to our results, we do not foresee $\bar{^3He}$ to be detected by AMS-02 in the next years. In turn, 1-4 GeV/n $\bar{d}$ are likely to be captured by AMS (RICH). To compare the differences between these diffusion models in secondary antinuclei fluxes, see Figure \ref{fig:Ratios_BF}. Excitingly, ALADInO is expected to be able to resolve and measure with high precision the secondary flux of $\bar{d}$ and $\bar{^3He}$, which will certainly allow us to improve our modelling of coalescence of antinucleons (See details  in Ref.~\cite{DeLaTorreLuque:2024htu}).
%The spectra of secondary antinuclei produced in the $\gamma$-inhomogeneous and inhomogeneous model do not exhibit visually noteworthy differences compared to the spectra shown below (see Fig. \ref{fig:Ratios}, \ref{fig:Ratios_BF} for discrepancies between different models in antinuclei predictions). 
%The best-fit inhomogeneous model, in turn, predicts a secondary antinuclei flux that is roughly 10$\%$ lower than the $\gamma$-inhomogeneous model, in accordance with what is shown in Figure \ref{fig:Ratios_BF}.

The production of antinuclei solely coming from DM annihilation follows a trend similar to that discussed for antiprotons in Section \ref{sec:DM}. 
In both NFW and c-NFW profiles (middle and bottom panels of Fig.~\ref{fig:ImhAN}, respectively), we find that the variations do not exceed 5$\%$ in the spectral peaks in any case, occasionally resulting in the antinuclei flux of the homogeneous model being slightly higher. This occurs in the case of m$_\chi\in$~[10-50]~GeV, where the flux in inhomogeneous diffusion is 3$\%$ lower than the homogeneous one. We observe a barely noticeable increase in flux in the c-NFW profile compared to the NFW profile, approximately in the range of 1-2$\%$. It is in the spectral peaks of m$_\chi\in$~[500-1000]~GeV where the best-fit inhomogeneous mdoel is larger over the homogeneous one, with an increase between models of about 2-3$\%$.

\iffalse
\begin{figure}[t!]
    \centering
    \includegraphics[width=0.47\textwidth]{Figures/Ad_DM.png}
    \includegraphics[width=0.47\textwidth]{Figures/AHe_DM.png}
    \caption{Predicted antinuclei flux ratio comparison from DM annihilation via $b\bar{b}$ for different WIMP masses. Solid line is the flux predicted by the inhomogeneous model. Dashed line is for homogeneous model.}
    \label{fig:AN_DM}
\end{figure}
\fi

\newpage
\bibliographystyle{apsrev4-1}
\bibliography{biblio}

%\FloatBarrier
%\textfloatsep 2.5cm

\end{document}